\documentclass[%
 aip,
 amsmath,amssymb,
 preprint,%
]{revtex4-1}

\usepackage{graphicx}
\usepackage{dcolumn}
\usepackage{bm,xcolor}

\usepackage[mathlines]{lineno}

\usepackage[utf8]{inputenc}
\usepackage[T1]{fontenc}
\usepackage{mathptmx}

\begin{document}


\title[Micromagnetic modeling of magnon coherent states in a nonuniform magnetic field]{Micromagnetic modeling of magnon coherent states in a nonuniform magnetic field}


\author{A.D. Belanovsky}
\email{abelanovsky@gmail.com}
\affiliation{Crocus Nanoelectronics, Volgogradskiy prospect 42 build. 5, premise 1, Moscow, 109316, Russia}

\author{Yu.M. Bunkov}
\affiliation{Vernadsky Crimean Federal University, Simferopol, 295007, Russia}
\affiliation{M-Granat, Russian  Quantum  Center, Skolkovo, Bolshoy Bulvar 30, bld. 1, Moscow, 121205, Russia}

\author{P.M. Vetoshko}
\affiliation{Vernadsky Crimean Federal University, Simferopol, 295007, Russia}
\affiliation{Kotelnikov Institute of radioengineering and electronics of RAS, Mokhovaya str., 11-7, Moscow 125009, Russia}


\date{\today}

\begin{abstract}
The study of the dynamics of magnetically ordered states in strong excitation through micromagnetic modeling has become relevant due to the observation of magnon Bose condensation. In particular, the question has arisen about the possibility of describing the coherent quantum state by the quasi-classical Landau-Lifshitz-Gilbert equations. We performed micromagnetic simulations of magnetization precession with a high angle of deviation in an out-of-plane nonuniform dc field. Our results confirm the formation of coherent magnon state under conditions of high excitation. This coherent state extends over long distances and described by a spatially inhomogeneous amplitude and a homogeneous precession phase.
\end{abstract}

\maketitle
\section{Introduction}
The magnetically ordered materials in a quantum approximation can be described by the ground state and by the gas of bosonic excitations - magnons. When magnon density is sufficiently high the magnons condense into a coherent quantum state  (the Bose - Einstein condensed state, mBEC).  During the thermal equilibrium, the magnon density is below the critical density of  mBEC, so it can be significantly enhanced by magnetic resonance. The first magnon BEC was observed in an antiferromagnetic superfluid $^3$He-B (see Refs. \onlinecite{Borovik-Romanov1984,Bunkov2008}) and revealed itself as a spontaneously self-organized phase-coherent precession of magnetization. Furthermore, the magnon supercurrent was discovered, which has all the properties specific to superfluidity and superconductivity \cite{Borovik-Romanov1989-2,Borovik-Romanov1987} such as long-distance magnetization supercurrent in the channel, phase-slippage at critical flow, the Josephson magnon effect, and magnon-current vortices. There have been several reviews of magnon supercurrent and mBEC investigations in antiferromagnetic (see Refs. \onlinecite{Bunkov2009,Bunkov2013}). 

Due to the interaction between magnons, the frequency of precession depends on their concentration, the so-called dynamic frequency shift. This shift is positive in the case of repulsive interaction.  The energy  of repulsive interaction provides potential in the form of a Mexican hat, which stabilizes the magnon superfluidity \cite{Bunkov2013}. The magnetic field spatial inhomogeneity forms a precession phase gradient, which excites the supercurrent of magnons.  The supercurrent forms a state with coherent precession by redistributing magnons until the dynamic frequency shift compensates for the inhomogeneity of an effective magnetic field, as demonstrated  by specially designed  direct experiments \cite{Borovik-Romanov1985}. This process is very similar to the Meissner phenomenon in superconductors \cite{Bunkov2020}. 

Though the experiments were performed on antiferromagnetic states of the superfluid ${}^3$He, its properties are not related directly to mass superfluidity and can be found in other magnetically ordered materials. The main difference between magnons in $^3$He-B and magnons in solid magnetics is is that the latter has a very long lifetime. However magnon superfluidity and coherent precession state have been observed in solid materials, including antiferromagnets with coupled nuclear-electron precession \cite{Bunkov2012,Abdurakhimov2018,Bunkov2019} and in out-of-plane magnetized YIG film \cite{Bunkov2021,Vetoshko2020} where the magnon interaction is repulsive. 

Furthermore, the magnon BEC state was observed in in-plane magnetized YIG film. In this case, the magnon interaction is attractive, and the energy minimum corresponds to spin waves with non-zero wavenumber\cite{Dzyapko2017}. Thus, the BEC state is unstable for magnons with wavenumber $k=0$. 
The formation of magnon BEC for these waves has been discussed since 2006 (see Refs. \onlinecite {Demokritov2006,Snoke2006}) and was observed directly in 2014 (see Ref. \onlinecite{Serga2014}). The recent review of mBEC experimental investigations in this configuration can be found in Ref. \onlinecite {Noack2021}.

The relation between spin quantum properties and its  description by the classical approximation was previously discussed in the 1960s. Macroscopic excitations in nonuniform external and internal magnetic fields were studied semiclassically by Schl\"{o}mann \cite{Schlomann1964} and Eshbach \cite{Eshbach1963}. In these works, a close analogy was noted between the propagation of exchange-dominated spin waves and the movement of a quantum mechanical particle in a potential well. Rezende and Zagury also showed the relationship between semiclassic spin waves and coherent quantum magnon states \cite{Rezende1969, Zagury1969}. Interest in these investigations has now been revived due to experimental observation of magnon BEC. Magnon BEC is quantum phenomenon. The direct application of quantum theory for studies of the spatial distribution of magnons is very complicated. However, the observed values of quantum processes can be described in a quasi-classical approximation. In this article, we used the quasi-classical Landau-Lifshitz-Gilbert equation (LLGE) to simulate the results of quantum processes for the formation of the coherent state of magnion.  We theoretically show that in out of plane magnetized YIG film with linear ramp external field $B_z(x)$, a magnon coherent state can be formed with $k=0$, and it has the same behavior as BEC in ${}^3$He-B and other known atomic condensates.

\section{Simulation details and results}

\begin{figure}
    \centering
    \includegraphics[]{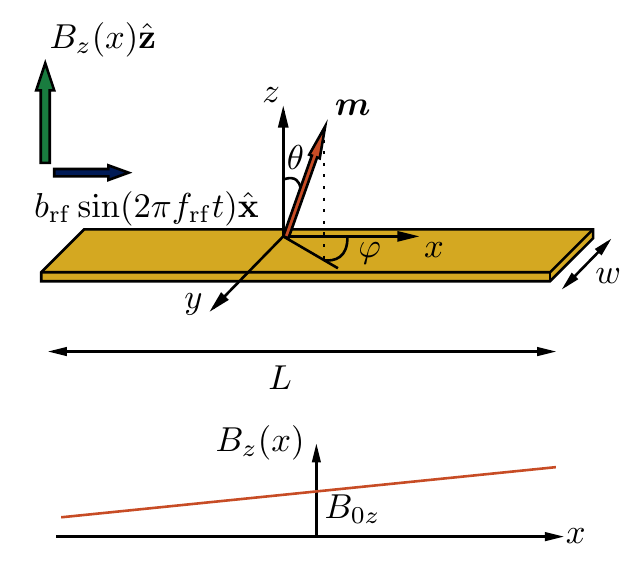}
    \caption{Sketch of the system studied. Rectangular YIG film with length $L = 100~\mu$m, width $w = 2~\mu$m, and thickness $h = 50$ nm. Spatially non-uniform external magnetic field was applied perpendicular to the film plane with an rf pumping field along the $x$-axis $\textbf{B}_{\mathrm{ext}} = (b_{\mathrm{rf}}\sin(2\pi f_{\mathrm{rf}}t),~0,~B_z(x))$. $B_z(x)$ has linear dependence on the spatial coordinate $x$. The unit vector of magnetization $\bm{m}$ rotates around the $z$-axis with precession amplitude $\theta$ and phase $\varphi$.}
    \label{fig1}
\end{figure}

In the present numerical study, we use the open-source finite-difference micromagnetic solver MuMax$^3$ (see Ref. \onlinecite{Vansteenkiste2014}). Calculations were done on the high performance platform ''Zhores'' based on SkolTech \cite{Zacharov2019}. We simulate magnetization dynamics at zero temperature in rectangular film with length $L = 100~\mu$m, width $w = 2~\mu$m, thickness $h = 50$ nm, and computational cell size $10\times10\times50$ nm$^3$ (see Fig. \ref{fig1}). To mimic a relatively large sample with uniform demagnetization field $B_{\mathrm{demag}z} = -\mu_0M_S$, periodic boundary conditions (PBC) with 10 repetitions were applied along the $x$- and $y$-axes. The material parameters such as Gilbert damping, saturation field and nonuniform exchange constant were chosen as follows \cite{Klingler2014}:
$$\alpha_G = 10^{-5},~\mu_0M_S = 165~\text{mT},~A = 3.49\times10^{-12}~\text{J/m}.$$
Crystallographic anisotropy has been neglected. Perpendicular to the film plane magnetic field as a function of spatial coordinate $x$ has linear dependence:
\begin{equation}
B_z(x) = B_{0z}\left(1 + \frac{\delta}{L}x\right).
\end{equation}
In our case, we chose $B_{0z} = 250$ mT and $\delta = 0.08$. An in-plane microwave field $b_{\mathrm{rf}}$ has orders of magnitude 25 - 200 $\mu$T and frequency $f_{\mathrm{rf}} = \gamma(B_{0z} - \mu_0M_S)=2.38$ GHz ($\gamma = 28$ GHz/T), which corresponds to the resonance frequency at $B_{0z}$. To achieve the steady-state, the simulation was run for more than 50 $\mu$s.

\begin{figure}
\centering
\includegraphics[]{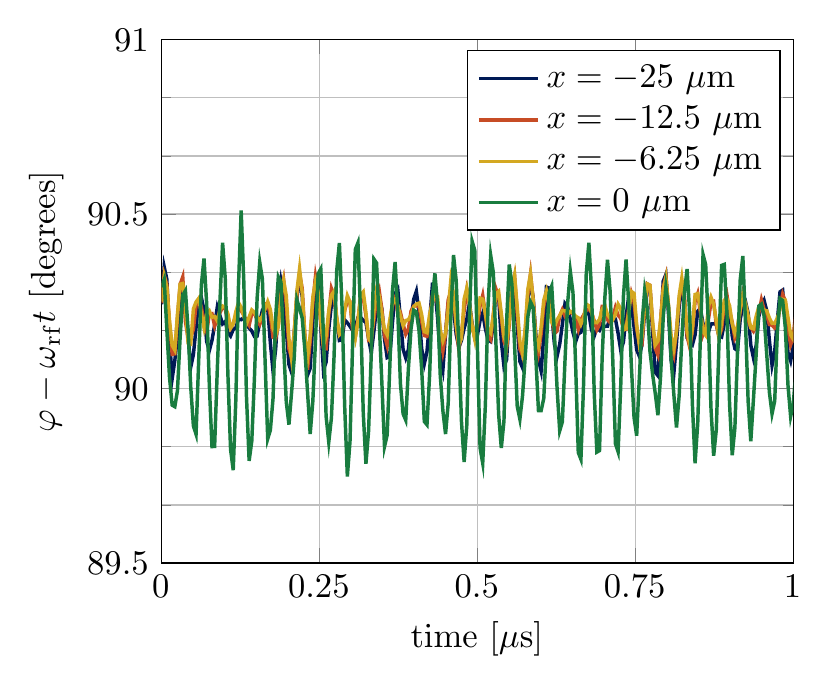}
\caption{Time evolution of the phase difference $\varphi-\omega_{\mathrm{rf}}t$ at $b_{\mathrm{rf}}=100~\mu$T in different sample $x$ locations ($y = 0$) after waiting for steady state.}
\label{fig2}
\end{figure}

For the microwave field with amplitude $b_{\mathrm{rf}}=100~\mu$T, the time evolution of the difference between the phase of the magnetization precession and the rf pumping field $\varphi - \omega_{\mathrm{rf}}t$ (here $\omega_{\mathrm{rf}} = 2\pi f_{\mathrm{rf}}$) at the stationary regime presented in Fig. \ref{fig2}. As can be seen, the phase difference at  steady-state in a different sample location has a very small deviation (less than $1^{\circ}$) from the average value during the time, thus one can say that averaged values $\langle\varphi(t) - \omega_{\mathrm{rf}}t\rangle\equiv const$ and $\langle\theta(t,x)\rangle\equiv\theta(x)$.

\begin{figure}
    \centering
    \includegraphics[scale=0.99]{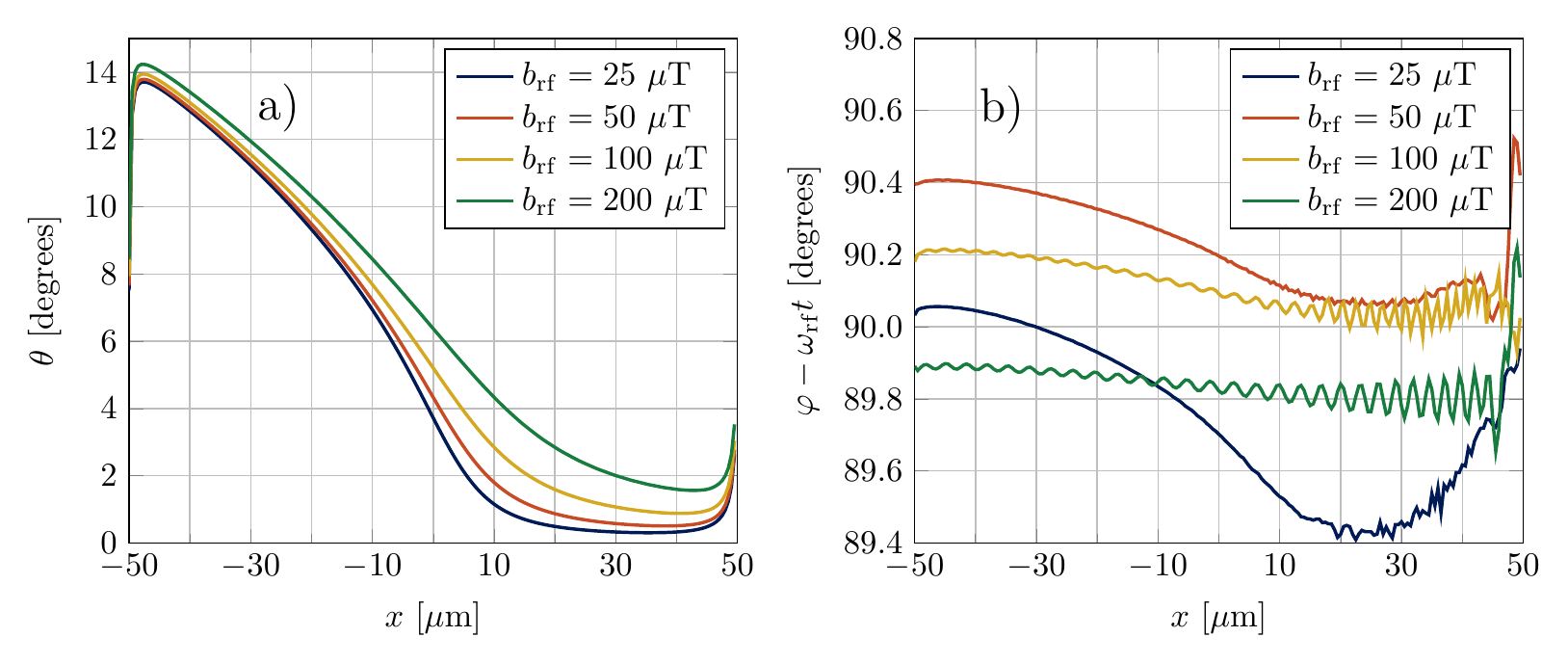}
    \caption{Averaged over 50 ns, (a) precession amplitude $\theta$ and (b) phase difference $\varphi-\omega_{\mathrm{rf}}t$ as a function of coordinate $x$ at $y=0$ for different rf pumping field magnitudes. Amplitude $\theta$ increases very little as $b_{\mathrm{rf}}$ increases, while $\varphi - \omega_{\mathrm{rf}}t$ stays steady at about 90 degrees along the whole sample, showing global coherence.}
    \label{fig3}
\end{figure}

\begin{figure}
\centering
\includegraphics[]{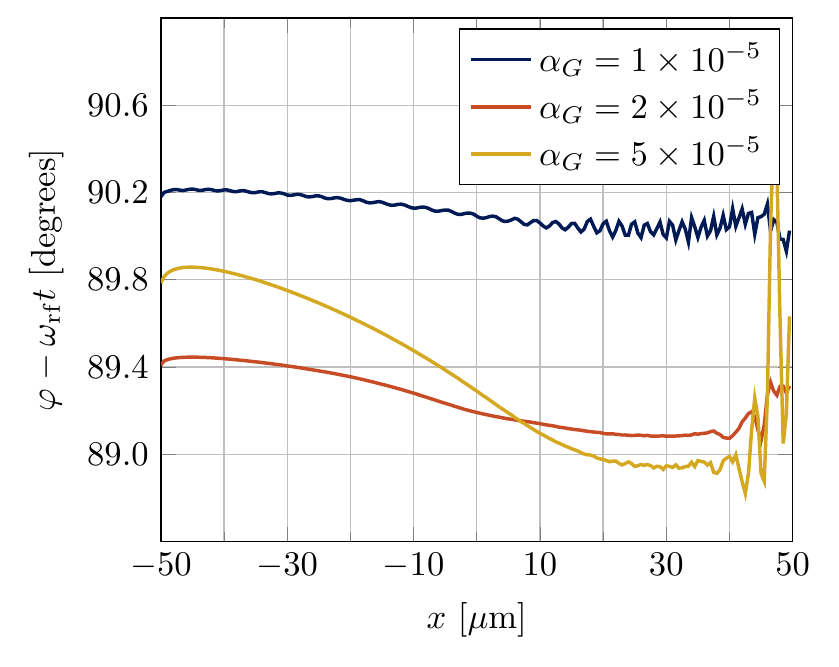}
\caption{Precession phase as a function of the $x$ coordinate for different Gilbert dampings for the rf pumping field $b_{\mathrm{rf}} = 100~\mu$T. Increasing of Gilbert damping can violate phase uniformity.}
\label{fig4}
\end{figure}

 The simulation results averaged over 50 ns at steady-state for other pumping fields presented in Fig. \ref{fig3}. As will be shown the stationary precession amplitude $\theta$ has the Airy asymptote at $x > 0$ (ignoring boundaries) and square root dependence $\theta\sim\sqrt{|x|}$ at $x < 0$. It is also interesting that in the low perpendicular field, the magnitude of $\theta$ has weak dependence on the rf field (less than 5\% with an 8-fold increase in pumping). This result differs from the classical Anderson-Suhl theory of nonlinear ferromagnetic resonance \cite{Anderson1955,Suhl1957} where the component $M_z$ has strong dependence on the rf field magnitude $b_{\mathrm{rf}}$. A similar effect (weak dependence of the precession amplitude on the microwave magnetic field) was obtained experimentally in YIG \cite{Vetoshko2020} and in superfluid ${}^3$He-B (see Ref. \onlinecite{Borovik-Romanov1989-1}). The phase difference $\varphi - \omega_{\mathrm{rf}}t$ is almost uniform along the whole sample and equals 90 degrees, which is striking evidence of  global coherence with $k = 0$. As can be seen from Fig. \ref{fig4}, low damping is crucial for achieving phase uniformity.

\section{Analytics}

In order to qualitatively understand the results obtained by micromagnetic modeling, we use the undamped Landau-Lifshitz equation (LLE) of motion for the unit vector of magnetization $\bm{m}$ within the limits of the small deviations.
\begin{equation}
\frac{\partial\bm{m}}{\partial t} = -\gamma \frac{2A}{M_S}(\bm{m}\times\Delta\bm{m}) - \gamma (\bm{m}\times\mathbf{B}_{\mathrm{demag}}) - \gamma (\bm{m}\times\mathbf{B}_{\mathrm{ext}}) \label{eq:LLE}
\end{equation}
Following the procedure described in Refs. \onlinecite{Corones1977, Kosevich1990} we rewrite Eq. \eqref{eq:LLE} for each vector $\bm{m}$ component, and after function $\Psi = m_x+im_y$ is introduced, one can obtain

\begin{equation}
  i\frac{\partial \Psi}{\partial t} + \gamma \frac{2A}{M_S}(\Delta m_z \Psi - m_z \Delta \Psi ) - \gamma \mu_0M_S m_z \Psi + \gamma B_z(x)\Psi = 0.
\end{equation}
Here we used the fact that in our case $\mathbf{B}_{\mathrm{demag}} = (0, 0, -\mu_0 M_S)$. For the small deviation of the vector $\bm{m}$ from the equilibrium direction? one can write $\theta \ll 1, ~ m_z \approx 1 - \frac{1}{2}|\Psi|^2$, which yields

\begin{equation}
i\hbar\frac{\partial \Psi}{\partial t} - \frac{\hbar^2}{2m} \frac{\partial^2 \Psi}{\partial x^2} + V(x)\Psi + g \Psi |\Psi|^2 = 0, \label{eq:NLSE}
\end{equation}
where $m = \hbar \frac{M_S}{4\gamma A},~V(x) = \gamma \hbar \left( B_{0z} - \mu_0 M_S  + B_{0z}\frac{\delta}{L}x\right),~\text{and}~g = \frac{1}{2} \gamma \hbar \mu_0 M_S> 0$.
Eq. \eqref{eq:NLSE} is a nonlinear Schr\"{o}dinger equation (NLSE) or Gross-Pitaevskii equation with repulsive interaction. Unlike Ref. \onlinecite{Schlomann1964}, where the similarity between Landau-Lifshitz and Schr\"{o}dinger equations was pointed out, in our case there is cubic nonlinearity which can give solutions other than the Airy function.


As has already been shown, in micromagnetic modeling, the wavenumber $k = 0$, therefore one can use the ansatz $\Psi = C(x)e^{i\omega t}$, where $\omega$ is the precession frequency. This ansatz leads us to the stationary equation
\begin{equation}
    \left[-\frac{\hbar^2}{2m}\frac{\partial^2}{\partial x^2} + Fx + g|C(x)|^2\right]C(x) = \gamma \hbar \left(\frac{\omega}{\gamma} - (B_{0z} - \mu_0M_S)\right)C(x), \label{eq:SNLSE}
\end{equation}
where $F = \gamma \hbar B_{0z}\frac{\delta}{L}$.

In our case $\omega = \gamma(B_{0z} - \mu_0M_S)$; thus the right hand side of Eq. \eqref{eq:SNLSE} equals to zero. Let us rescale Eq. \eqref{eq:SNLSE} according to $\xi = x/\beta$, where $\beta = \displaystyle \left(\frac{\hbar^2}{2mF}\right)^{1/3} = \left( \frac{2AL}{M_SB_{z0}\delta} \right)^{1/3}$ and $u(x) = \displaystyle C(x)\sqrt{\frac{g}{F\beta}} = C(x)\sqrt{\frac{\mu_0M_SL}{2B_{0z}\delta\beta}}$. Hence, we obtain
\begin{equation}
    u'' = \xi u + u^3. \label{eq:Painleve}
\end{equation}
Eq. \eqref{eq:Painleve} is a so-called Painlev\'e equation of the second type, which has been well investigated by several mathematicians (see, for example Ref. \onlinecite{Rosales1978}). It is also studied in the field of the Bose–Einstein condensate \cite{Lundh1997, Dalfovo1996}. In general, as mentioned in Ref. \onlinecite{Tuszynski2001} Eq. \eqref{eq:Painleve} has only one monotonic solution, and the rest are oscillatory. 

We will focus on the monotonic solution since it represents the ground state of the system. We divide the solution into two parts: for $\xi > 0$ we have the Airy asymptote, and for $\xi < 0$, $u = \sqrt{-\xi}$, which corresponds to Thomas-Fermi (TF) approximation. This solution is well consistent with the formation of a magnon coherent state in an inhomogeneous magnetic field. Any deviation from this state creates a spin superfluidity, which restores this solution. 

\begin{figure}
\centering
\includegraphics[]{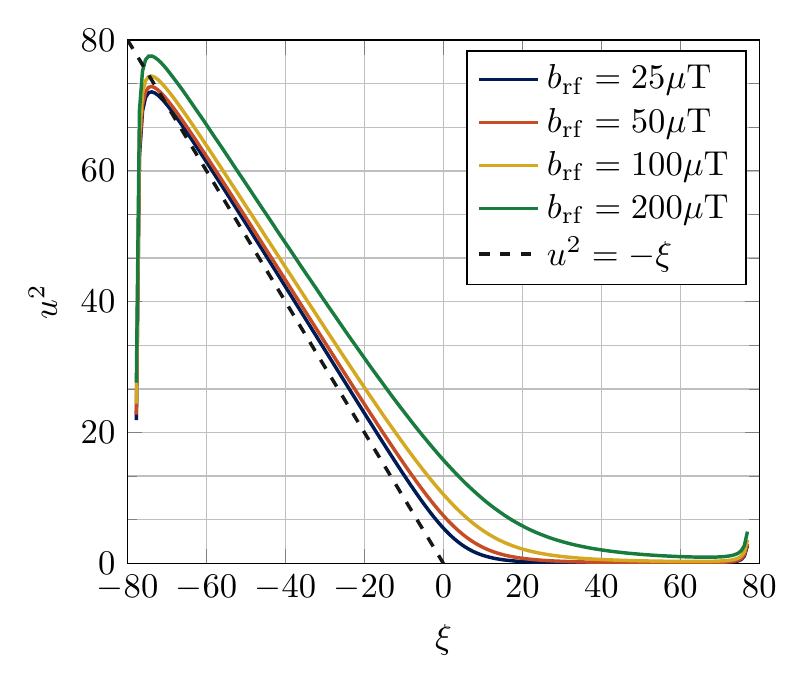}
\caption{Solid lines are rescaled micromagnetic simulations results; dashed line is analytical approximation based on Eq.
\eqref{eq:Painleve}.}
\label{fig5}
\end{figure}

The micromagnetic modeling and the TF solution are compared in Fig. \ref{fig5}. As can be seen, a full-scale simulation gives slightly higher values, which can be explained by the non-zero $x$-component of the magnetic field. However, the results coincide well with each other.

\begin{figure}
    \centering
    \includegraphics[]{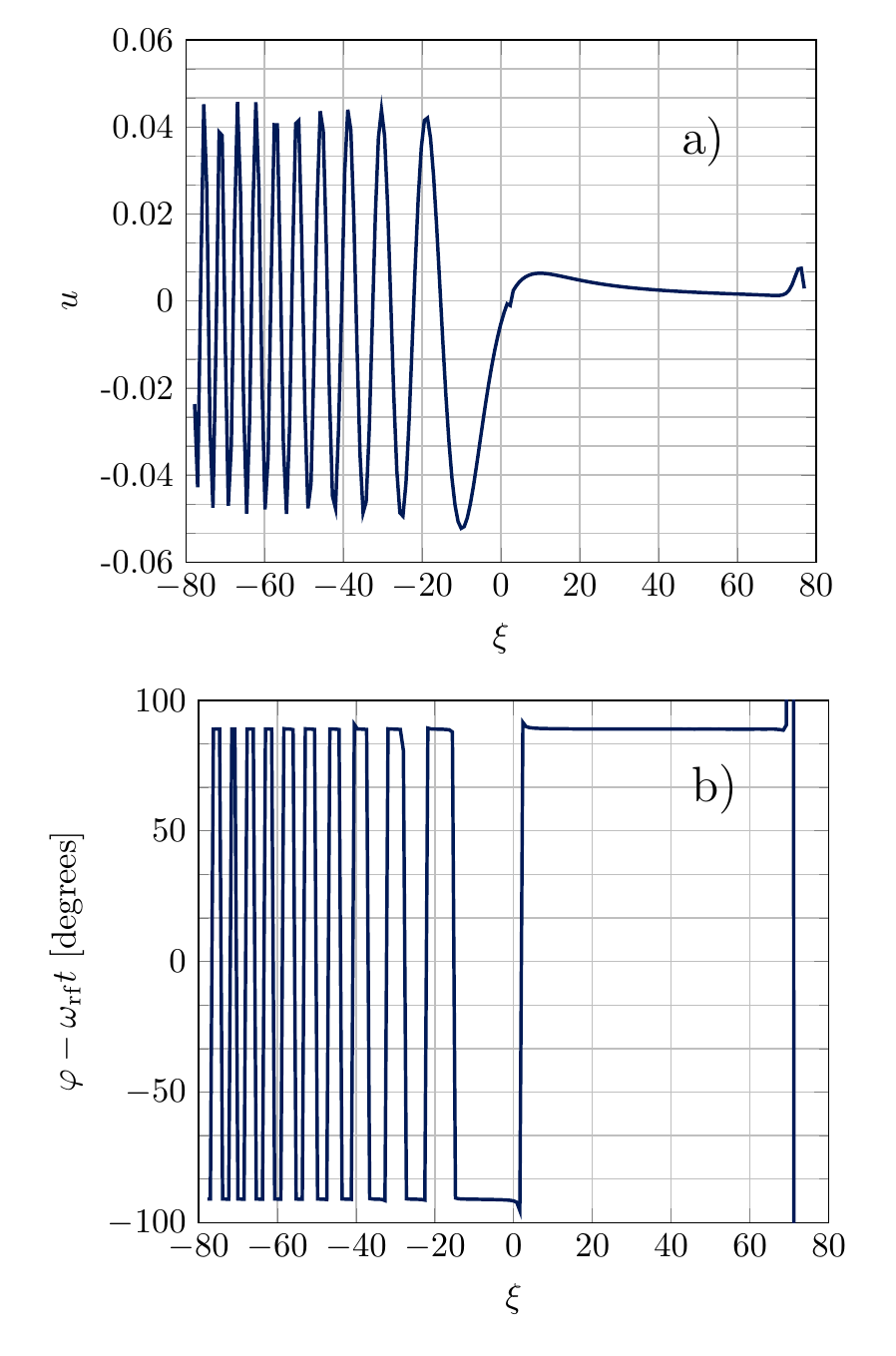}
    \caption{Rescaled micromagnetic simulation results for $b_{\mathrm{rf}} = 0.25~\mu$T and $\alpha_G=10^{-6}$. (a) Dimensionless amplitude $u$ and (b) phase difference as a function of dimensionless coordinate $\xi$.}
    \label{fig6}
\end{figure}

The oscillatory solution can be also found in micromagnetic simulations in small microwave fields.  However, in this case, due to relaxation, the phase is not uniform along the sample even at very small $\alpha_G = 10^{-6}$ (see Fig. \ref{fig6}).

\section{Conclusions}
By means of micromagnetic simulations, we have demonstrated the possibility of modeling the formation of the magnon coherence state in long YIG film under linear ramp perpendicular external field and an in-plane rf pumping field. We have shown that the precession amplitude does not depend on the magnitude of the rf field. The difference between the oscillation phase and the rf phase $\varphi - \omega_{\mathrm{rf}}t$ is very small; there is little deviation from 90 degrees along the whole sample. Our analytical approach based on the transition from LLE to NLSE has shown complete correspondence with atomic, liquid ${}^3$He-B or other types of BEC. 

We can also apply our modeling to the case of limited geometry, such as a thin disk, in which the demagnetization field has a nonlinear distribution. These calculations will be published elsewhere. The spatial distribution of magnetization in the form of magnon BEC in an inhomogeneous magnetic field was recently observed by magneto-optical methods and will be published in the near future.

From an application point of view, our results described in this article open the way to calculating specific properties for magnonic quantum calculations \cite{Bunkov2020-2} as well as for magnetic field sensors. In contrast with quasi-static magnetization rotation \cite{Koshev2021} the coherence states can drastically increase sensor dynamic range down to the quantum limit \cite{Kimball2016}.

\begin{acknowledgments}
This work was made in collaboration with the experimentalists from the Crimean Federal University and the Russian Quantum Center. It was financially supported by the Ministry of Science and Higher Education of the Russian Federation (Megagrant project no. 075-15-2019-1934).

\end{acknowledgments}


\bibliography{aipsamp}

 \newpage

\end{document}